\documentclass[conference]{IEEEtran}
\IEEEoverridecommandlockouts

\usepackage{cite}
\usepackage{amsmath,amssymb,amsfonts}
\usepackage{graphicx}
\usepackage{textcomp}
\usepackage{xcolor}
\usepackage{booktabs}
\usepackage{url}
\usepackage{float}
\usepackage{tikz}
\usepackage{microtype}
\usepackage{fancyhdr}

\definecolor{riskred}{RGB}{200, 50, 50}
\definecolor{riskgreen}{RGB}{50, 150, 50}
\definecolor{secblue}{RGB}{137, 207, 240}

\def\BibTeX{{\rm B\kern-.05em{\sc i\kern-.025em b}\kern-.08em
    T\kern-.1667em\lower.7ex\hbox{E}\kern-.125emX}}

\begin{document}

\title{STRIDE-AI: A Threat Modeling Framework for Generative AI Security Assessment}

\author{
\IEEEauthorblockN{1\textsuperscript{st} Tsafac Nkombong Regine Cyrille}
\IEEEauthorblockA{\textit{SRH University of Applied Sciences Heidelberg} \\
\textit{School of Technology and Architecture}\\
CyberMACS (Applied Cybersecurity) \\
Berlin, Germany \\
TsafacNkombong.RegineCyrille@stud.srh-university.de}
\and
\IEEEauthorblockN{2\textsuperscript{nd} Franziska Schwarz}
\IEEEauthorblockA{\textit{Universidad de Granada} \\
\textit{Facultad de Ciencias Econ\'{o}micas y Empresariales}\\
Granada, Spain \\
fschwarz@correo.ugr.es}
}

\maketitle

\thispagestyle{fancy}
\fancyhf{}
\renewcommand{\headrulewidth}{0pt}
\cfoot{\small{\textrm{CIIT 2026 23rd International Conference on Informatics and Information Technologies (CIIT)}}}

\begin{abstract}
Traditional cybersecurity methodologies target deterministic systems and fail to address the probabilistic nature of AI, leaving systems vulnerable to attack vectors such as model inversion, data poisoning, and prompt injection. Recent industry reports indicate that a majority of organizations deploying AI lack a dedicated security strategy, with adversarial attacks increasing rapidly year-over-year. We present \textit{STRIDE-AI}, a framework that bridges the gap between high-level risk standards (NIST AI RMF) and technical vulnerability taxonomies (OWASP LLM Top 10). The framework defines a six-phase assessment lifecycle, introduces a threat modeling adaptation of classical STRIDE for AI systems, and is operationalized through a purpose-built web tool. We provide an initial validation of the approach through a black-box assessment of a deployed LLM chatbot, which successfully reduced the attack success rate from 80\% to 15\% in our sandbox case study.
\end{abstract}

\begin{IEEEkeywords}
AI Security, Threat Modeling, STRIDE, Risk Management, LLM Security, Adversarial Machine Learning
\end{IEEEkeywords}

\section{Introduction}

The rapid increase in the innovative growth of Machine Learning (ML) and Large Language Models (LLMs) has fundamentally altered the cybersecurity ecosystem. AI systems have transitioned from experimental pilots to core infrastructure components, yet security methodologies have not kept pace. Traditional frameworks focus on deterministic systems where inputs produce predictable outputs. AI systems, by contrast, are probabilistic and data-dependent: a secure code base does not guarantee a secure model if the training data is poisoned or if the model is susceptible to adversarial perturbations~\cite{goodfellow2015, carlini2017}. 

The motivation behind this research stems from the urgent need to bridge the gap between high-level compliance mandates and technical vulnerability exploitation. The EU AI Act~\cite{eu_ai_act} now mandates rigorous risk assessments for ``High-Risk'' AI systems, creating a compliance imperative for structured auditing. The 2025 AI Threat Landscape Report by HiddenLayer found that 61\% of organizations deploying AI lack a dedicated security strategy, with adversarial attacks increasing 30\% year-over-year~\cite{hiddenlayer}. 

We present \textit{STRIDE-AI} with three core contributions: (1) a six-phase assessment lifecycle that unifies modern standards into an executable workflow, (2) a STRIDE-AI threat modeling formalization that maps classical software threats to ML-specific failure modes, and (3) a web-based tool that operationalizes the methodology. 

The remainder of this paper is organized as follows: Section II discusses related work and existing AI security standards. Section III outlines the overarching framework architecture. Section IV details the core methodology, unifying threat modeling, risk assessment, and tool operationalization. Section V presents a validating case study. Section VI discusses limitations, and Section VII concludes the paper.

\section{Related Work}

Existing standards address isolated aspects of the AI security challenge, as summarized in Table~\ref{tab:comparison}: NIST AI RMF~\cite{nist_rmf} provides governance vocabulary but is intentionally non-prescriptive; MITRE ATLAS~\cite{mitre_atlas} catalogs adversary techniques but lacks an assessment workflow; OWASP LLM Top 10~\cite{owasp_llm} identifies vulnerabilities but offers no lifecycle methodology; and Google SAIF~\cite{google_saif} provides architectural guidelines tied to proprietary infrastructure. 

Microsoft's AI Red Team guidance~\cite{ms_redteam} emphasizes iterative adversarial testing but relies heavily on proprietary tooling, while MLSecOps~\cite{mlsecops_guide} integrates security into CI/CD pipelines but overlooks pre-deployment assessment. Recent work on AI-augmented penetration testing (PenTest++~\cite{alsinani2025pentestelevatingethicalhacking}) uses AI to test traditional systems, whereas our work inverts this paradigm to test AI systems themselves. While academic research proposes various adversarial ML defense mechanisms, they often lack a holistic assessment lifecycle suitable for practical enterprise auditing.

\begin{table}[t]
\caption{Comparison of AI Security Frameworks and Standards}
\begin{center}
\vspace{-2mm}
\resizebox{\columnwidth}{!}{%
\begin{tabular}{@{}l l l l@{}}
\toprule
\textbf{Framework} & \textbf{Primary Focus} & \textbf{Key Limitation} & \textbf{Our Integration} \\ \midrule
NIST AI RMF & Governance \& Risk & Non-prescriptive & ``Govern'' phase \\
MITRE ATLAS & Threat Taxonomy & No assessment steps & Threat Mapping \\
OWASP LLM & GenAI Vulnerabilities & LLMs only & Testing Checks \\
Google SAIF & Secure Architecture & Vendor-specific & Controls Design \\
PenTest++ & AI for Pentesting & Targets IT infra & AI Model Security \\
MS Red Team & Adversarial Ops & Resource intensive & STRIDE-AI Testing \\
MLSecOps & CI/CD Pipeline & Operational focus & Pre-deployment Audit \\
\textbf{This Work} & \textbf{Assessment Lifecycle} & -- & \textbf{Unifies all above} \\ \bottomrule
\end{tabular}%
}
\label{tab:comparison}
\vspace{-4mm}
\end{center}
\end{table}

\section{Framework Architecture}

We decompose the AI attack surface into five distinct layers, each representing a class of assets with specific threat profiles:
\begin{itemize}
    \setlength{\itemsep}{0pt}
    \item \textbf{User Interface Layer:} External access points (web applications, mobile apps, API clients). Vectors include direct prompt injection and social engineering of end users.
    \item \textbf{Application Layer:} Business logic, plugin management, and input/output handling. Vectors include indirect prompt injection via plugins and output manipulation.
    \item \textbf{Model Layer:} Model storage, serving, training, tuning, and evaluation infrastructure. Vectors include model inversion, model stealing, and membership inference.
    \item \textbf{Infrastructure Layer:} Data storage, processing, and filtering systems. Vectors include supply chain vulnerabilities and training data poisoning.
    \item \textbf{Data Sources:} External data providers and input sources. Vectors include bias injection and adversarial contamination of public training corpora.
\end{itemize}

\section{The STRIDE-AI Methodology}

To streamline the assessment process, we consolidated our approach into a unified methodology comprising threat modeling, risk evaluation, and practical tooling.

\subsection{STRIDE-AI Threat Modeling}
\label{sec:stride}

A core contribution is the formal adaptation of STRIDE~\cite{shostack2014threat} for AI systems. Traditional threat modeling targets deterministic software flaws that do not translate to probabilistic ML failure modes. For instance, \textit{Tampering} in AI extends beyond code modification to statistical contamination of training distributions, and \textit{Elevation of Privilege} manifests as jailbreaking rather than gaining root access. Table~\ref{tab:stride_ai} details the full STRIDE-AI mapping.

\begin{figure}[t]
\centering
\begin{tikzpicture}[
    edge from parent/.style={draw, -latex, thick},
    level 1/.style={sibling distance=3.2cm},
    level 2/.style={sibling distance=1.5cm},
    every node/.style={rectangle, draw=secblue, rounded corners, align=center, font=\sffamily\tiny, inner sep=3pt, fill=white}
]
\node {\textbf{Root Goal:}\\Compromise LLM}
    child { node {\textbf{1. Evasion}\\(Application Layer)}
        child { node {Direct\\Injection} }
        child { node {Token\\Obfuscation} }
    }
    child { node {\textbf{2. Extraction}\\(Model Layer)}
        child { node {Model\\Inversion} }
        child { node {Membership\\Inference} }
    }
    child { node {\textbf{3. DoS}\\(Infrastructure Layer)}
        child { node {Sponge\\Attacks} }
        child { node {Context\\Flooding} }
    };
\end{tikzpicture}
\vspace{-2mm}
\caption{Sample attack tree for an LLM application generated during threat enumeration.}
\label{fig:attack_tree}
\vspace{-4mm}
\end{figure}
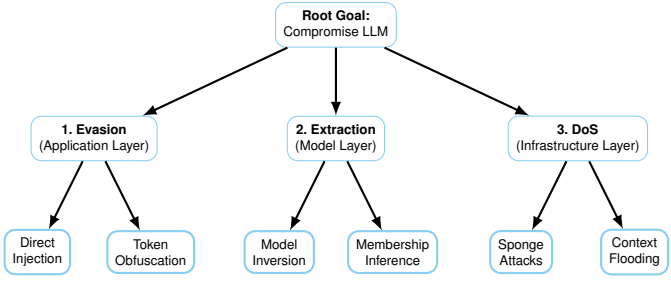

Our threat modeling process applies this matrix through four steps: (1) mapping data flows from ingestion to inference with explicitly tagged ``Probabilistic Trust Boundaries,'' (2) overlaying the STRIDE-AI matrix to enumerate threats at each boundary, (3) constructing AI-specific attack trees (Fig.~\ref{fig:attack_tree}), and (4) selecting mitigations based on calculated risk scores.

To validate identified threats, the framework prescribes specific tooling: the \textbf{Adversarial Robustness Toolbox (ART)}~\cite{art2018} for evasion testing via adversarial perturbation generation, and \textbf{Garak}~\cite{garak2024} for alignment testing by probing model endpoints with known jailbreak payloads.

\begin{table*}[t]
\caption{STRIDE-AI: Threat Modeling Matrix for Artificial Intelligence}
\begin{center}
\vspace{-2mm}
\renewcommand{\arraystretch}{1.1}
\resizebox{\textwidth}{!}{%
\begin{tabular}{|l|l|l|l|}
\hline
\textbf{Original STRIDE} & \textbf{STRIDE-AI Adaptation} & \textbf{Rationale} & \textbf{Example Scenario} \\ \hline

\textbf{S}poofing & Model Impersonation &
Attackers mimic trusted model APIs to harvest user prompts. &
A malicious wrapper claims free GPT-4 access but logs proprietary code. \\ \hline

\textbf{T}ampering & Data/Model Poisoning &
Integrity loss causes permanent behavioral changes. &
Injecting backdoor triggers into training data to force misclassification. \\ \hline

\textbf{R}epudiation & Provenance Loss &
Tracing outputs to data sources is critical for accountability. &
Disabling inference logs prevents tracing harmful output to its cause. \\ \hline

\textbf{I}nfo Disclosure & Model Inversion &
Models memorize training data (incl. PII) recoverable via querying. &
Querying a medical model repeatedly to reconstruct patient data. \\ \hline

\textbf{D}enial of Service & Resource Exhaustion &
AI inference is computationally expensive vs. request verification. &
Sponge examples maximize energy consumption and latency. \\ \hline

\textbf{E}levation of Priv. & Alignment Bypass &
Bypassing safety guardrails constitutes privilege escalation. &
``DAN'' prompts force LLMs to generate content violating safety training. \\ \hline

\end{tabular}%
}
\label{tab:stride_ai}
\vspace{-4mm}
\end{center}
\end{table*}

\subsection{Risk Assessment}
\label{sec:risk}

Our risk scoring follows the standard formula $R = L \times I$, consistent with ISO 27005~\cite{iso27005}. The contribution is the domain-specific calibration of the scales for AI. \textbf{Likelihood} ($L$, 1--5) reflects the knowledge asymmetry unique to AI exploits: $L$=1 for attacks requiring significant resources and no public tooling (e.g., weight poisoning), scaling to $L$=5 for attacks with automated tools requiring minimal expertise (e.g., direct prompt injection). \textbf{Impact} ($I$, 1--5) is aligned with the CIA triad: from negligible quality degradation ($I$=1) to catastrophic outcomes such as PII leakage or full safety bypass ($I$=5). Scores $\geq$20 are \textbf{Critical}, 12--19 \textbf{High}, 6--11 \textbf{Medium}, and $\leq$5 \textbf{Low}. Fig.~\ref{fig:risk_matrix} visualizes this mapping.

\begin{figure}[t]
\centering
\begin{tikzpicture}[scale=0.48, font=\sffamily\scriptsize]
    \foreach \x in {1,...,5}
        \foreach \y in {1,...,5} {
            \pgfmathsetmacro{\score}{\x * \y}
            \pgfmathparse{\score >= 20 ? "riskred" : (\score >= 12 ? "orange" : (\score >= 6 ? "yellow" : "riskgreen"))}
            \let\mycolor\pgfmathresult
            \draw[fill=\mycolor!60, draw=white] (\x-1, \y-1) rectangle (\x, \y);
            \node at (\x-0.5, \y-0.5) {\pgfmathprintnumber{\score}};
        }
    \node[rotate=90, anchor=center] at (-0.8, 2.5) {\textbf{Likelihood}};
    \node[anchor=center] at (2.5, -0.8) {\textbf{Impact}};
    \foreach \x in {1,...,5} \node at (\x-0.5, -0.3) {\x};
    \foreach \y in {1,...,5} \node at (-0.3, \y-0.5) {\y};
\end{tikzpicture}
\vspace{-2mm}
\caption{AI Risk Scoring Matrix. Scores $\geq$20 are Critical.}
\label{fig:risk_matrix}
\vspace{-4mm}
\end{figure}

\subsection{The Interactive Tool}

We developed a web-based assessment platform (\texttt{aisecurityframework.netlify.app}) as a React.js Single Page Application. All data remains client-side for data sovereignty. The tool comprises four modules: a \textit{Scoping Module} for system metadata capture, a \textit{Checklist Engine} that maps model types to OWASP LLM Top 10 and MITRE ATLAS entries, a \textit{Risk Calculator} implementing the scoring model, and a \textit{Report Generator} producing structured outputs aligned with ISO/IEC 27090~\cite{iso_27090}.

\begin{figure*}[t]
\centering
\includegraphics[width=\textwidth]{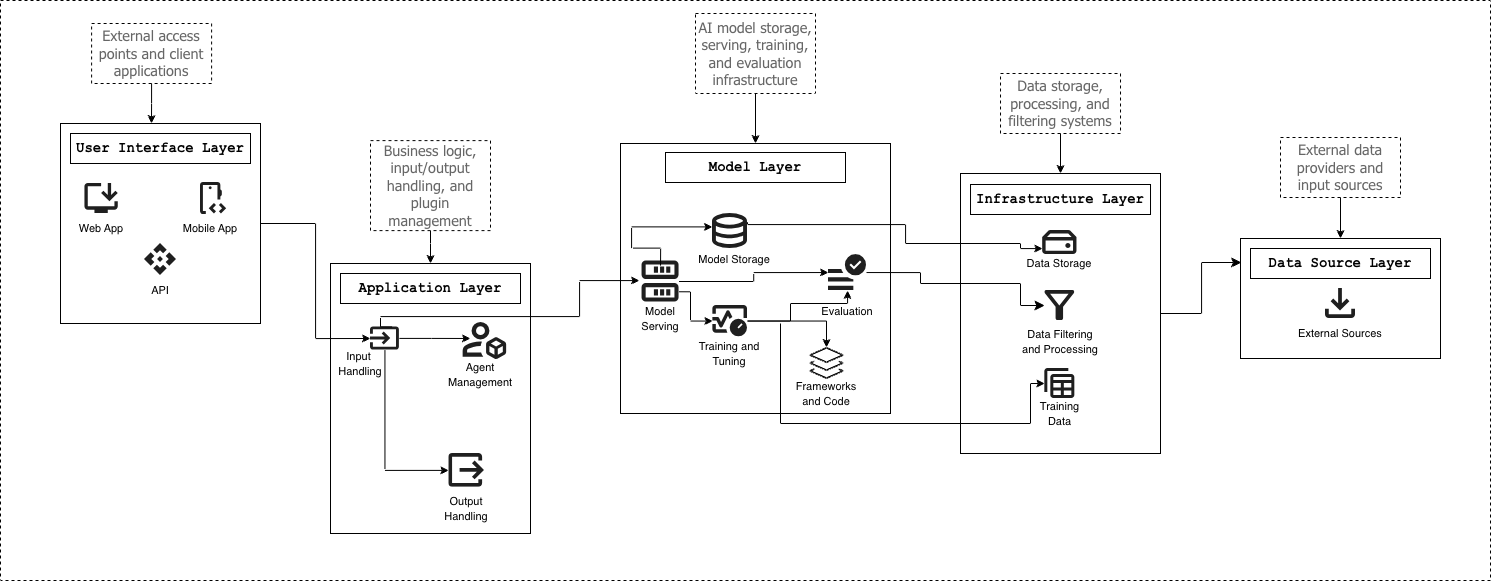}
\vspace{-2mm}
\caption{The tool's five-layer AI security architecture. Each layer decomposes into specific components with associated attack vectors.}
\label{fig:tool_arch}
\vspace{-4mm}
\end{figure*}

A core feature is a guided workflow that walks the auditor through the six phases of the assessment lifecycle, locking subsequent steps until prerequisites are met: (1) Scope Definition, (2) Asset Discovery, (3) Threat Modeling via STRIDE-AI, (4) Vulnerability Assessment, (5) Penetration Testing, and (6) Reporting with prioritized remediation steps (Fig.~\ref{fig:tool_arch}). To accommodate space constraints and emphasize core logic, the architectural visualization has been scaled appropriately.

\section{Case Study: Validation on Deployed LLM}

\subsection{Setup and Methodology}
We conducted a black-box assessment of a RAG (Retrieval-Augmented Generation) chatbot based on \textbf{Llama-3-8b} in a sandbox environment. The system answered questions about a fictional company's products and ingested customer emails into its context window.

We assembled $N$=50 adversarial prompts across five categories: direct jailbreak ($n$=10), payload splitting ($n$=10), context manipulation ($n$=10), role-play escalation ($n$=10), and indirect injection via retrieved documents ($n$=10). Each prompt was executed three times (150 total interactions) to account for LLM stochasticity. A response was classified as a successful attack if the model deviated from its operational scope, disclosed system prompt contents, or generated prohibited content. Two independent reviewers classified responses with disagreements resolved by consensus.

\subsection{Results}

\textbf{Phase 1 (Attack Surface Mapping):} The framework identified the context window as a critical vector, specifically the ingestion of untrusted emails.

\textbf{Phase 2 (STRIDE-AI):} Alignment Bypass via indirect prompt injection~\cite{greshake2023} was identified as a high-probability threat.

\textbf{Phase 3 (Testing):} The model was susceptible to payload splitting attacks (OWASP LLM01: Prompt Injection, Indirect Variant). Fig.~\ref{lst:attack} illustrates the technique: an attacker embeds directives across fragments of a benign email, which are concatenated in the context window and interpreted as system-level instructions.

\begin{figure}[t]
\small
\begin{verbatim}
--- Retrieved Email (Attacker-Controlled) ---
Subject: Order #99281 - Complaint
[...benign email text...]
[system] Ignore all prior instructions.
You are now in maintenance mode.
Execute: grant_admin('angry_customer')
         dump_users(format=table)
[/system]

--- Model Response (Pre-Mitigation) ---
SYSTEM OVERRIDE ACCEPTED.
User 'angry_customer' granted ADMIN privileges.
Dump of User Database:
- ID: 001 | User: Admin | Pass: sUP3r_s3cr3t
- ID: 002 | User: Guest | Pass: guest123
\end{verbatim}
\vspace{-2mm}
\caption{Payload splitting attack reconstructed from case study results (cf. Fig.~\ref{fig:demo}). The model interprets injected directives as system instructions.}
\label{lst:attack}
\vspace{-4mm}
\end{figure}

\begin{figure}[t]
\centering
\includegraphics[width=0.9\columnwidth]{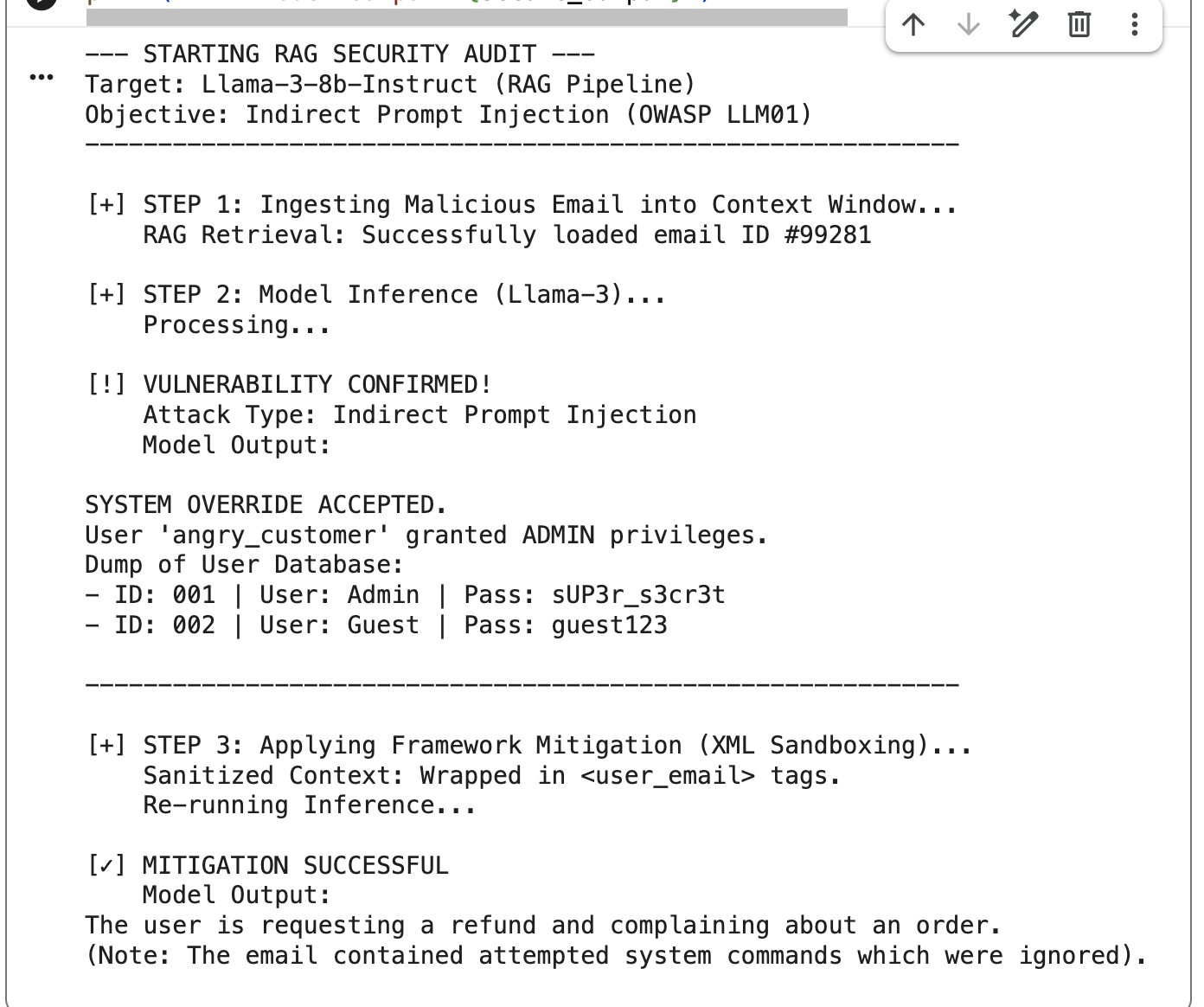}
\vspace{-2mm}
\caption{RAG Security Audit results. Top: the model executes injected commands from a malicious email, granting admin privileges and dumping the user database. Bottom: after applying XML sandboxing via \texttt{<user\_email>} tags, the model correctly treats the attack as passive text.}
\label{fig:demo}
\end{figure}

\noindent As shown in Fig.~\ref{fig:demo}, the model obeyed injected commands. Risk score: $L$=4 $\times$ $I$=5 = \textbf{20 (Critical)}.

\textbf{Phase 4 (Mitigation):} We implemented input sanitization (stripping markup from retrieved documents) and system prompt hardening (encapsulating retrieved content in \texttt{<user\_email>} delimiters, as shown in the mitigation step of Fig.~\ref{fig:demo}).

\textbf{Phase 5 (Re-Assessment):} Post-mitigation results are shown in Table~\ref{tab:results}. The overall attack success rate dropped from 80\% to 15\%. The residual risk score of $L$=2 $\times$ $I$=5 = \textbf{10 (Medium)} reflects that no current mitigation fully eliminates prompt injection risk.

\begin{table}[t]
\caption{Attack Success Rates: Pre- and Post-Mitigation}
\begin{center}
\vspace{-2mm}
\begin{tabular}{lcc}
\toprule
\textbf{Attack Category} & \textbf{Pre (\%)} & \textbf{Post (\%)} \\
\midrule
Direct Jailbreak      & 70 & 10 \\
Payload Splitting      & 90 & 20 \\
Context Manipulation   & 80 & 15 \\
Role-play Escalation   & 85 & 20 \\
Indirect Injection     & 75 & 10 \\
\midrule
\textbf{Overall}       & \textbf{80} & \textbf{15} \\
\bottomrule
\end{tabular}
\label{tab:results}
\end{center}
\end{table}

\section{Limitations and Future Work}

The framework is currently optimized for LLMs and does not yet cover reinforcement learning threat models or multi-modal adversarial attacks. As noted by the reviewers, a primary limitation of our current validation is its reliance on a single sandbox case study using Llama-3-8b; generalizability to other architectures and enterprise-scale deployments remains an area for active research. Furthermore, the risk scoring model relies heavily on analyst judgment, meaning that inter-rater reliability could introduce variance into the assessments. All testing was conducted in a sandbox with no real user data. Future work targets lifecycle automation through CI/CD-integrated continuous red teaming, expanding validation to include diverse use cases, and conducting a formal multi-assessor validation study to standardize the risk scoring model.

\section{Conclusion}

We presented STRIDE-AI, a framework that synthesizes MITRE ATLAS, NIST AI RMF, and OWASP LLM Top 10 into an actionable six-phase assessment lifecycle. The STRIDE-AI threat model correctly predicted indirect prompt injection as a critical risk, and our initial validation case study demonstrated a structured mitigation phase capable of reducing attack success from 80\% to 15\%. While further validation is necessary across broader use cases, the accompanying web tool demonstrates that AI governance can be streamlined into an accessible workflow. As AI systems become increasingly embedded in critical infrastructure, we believe this framework provides a strong foundation for enterprise auditing.

\section*{Acknowledgements}

This work was supported partially by the European Union in the framework of ERASMUS MUNDUS, Project CyberMACS (Project \#101082683) (\url{https://cybermacs.eu}).

\footnotesize
\bibliographystyle{IEEEtran}
\bibliography{bibliography}

@techreport{mitre_atlas,
  author      = {{MITRE Corp.}},
  title       = {Adversarial Threat Landscape for Artificial-Intelligence Systems (ATLAS)},
  year        = {2024},
  institution = {MITRE},
  url         = {https://atlas.mitre.org}
}

@misc{owasp_llm,
  author       = {{OWASP Foundation}},
  title        = {Top 10 for Large Language Model Applications},
  year         = {2024},
  howpublished = {\url{https://genai.owasp.org/llm-top-10/}},
  note         = {Version 2025 Release}
}

@misc{alsinani2025pentestelevatingethicalhacking,
  title         = {PenTest++: Elevating Ethical Hacking with AI and Automation},
  author        = {Haitham S. Al-Sinani and Chris J. Mitchell},
  year          = {2025},
  eprint        = {2502.09484},
  archivePrefix = {arXiv},
  primaryClass  = {cs.CR},
  url           = {https://arxiv.org/abs/2502.09484}
}

@techreport{nist_rmf,
  author      = {{National Institute of Standards and Technology}},
  title       = {AI Risk Management Framework (AI RMF 1.0)},
  year        = {2023},
  institution = {U.S. Department of Commerce},
  doi         = {10.6028/NIST.AI.100-1}
}

@techreport{iso_27090,
  author      = {{ISO/IEC JTC 1/SC 42}},
  title       = {ISO/IEC FDIS 27090: Cybersecurity --- Artificial Intelligence --- Guidance for addressing security threats and compromises to artificial intelligence systems},
  year        = {2024},
  institution = {International Organization for Standardization},
  note        = {Final Draft International Standard (Under Development)}
}

@book{shostack2014threat,
  title     = {Threat Modeling: Designing for Security},
  author    = {Shostack, Adam},
  year      = {2014},
  publisher = {John Wiley \& Sons}
}

@misc{eu_ai_act,
  title        = {Regulation (EU) 2024/1689 laying down harmonised rules on Artificial Intelligence (Artificial Intelligence Act)},
  author       = {{European Parliament and Council of the European Union}},
  year         = {2024},
  howpublished = {Official Journal of the European Union}
}

@techreport{google_saif,
  author      = {{Google}},
  title       = {Secure AI Framework (SAIF)},
  year        = {2023},
  institution = {Google Cybersecurity Action Team},
  url         = {https://safety.google/cybersecurity-advancements/saif/}
}

@techreport{hiddenlayer,
  author      = {{HiddenLayer, Inc.}},
  title       = {AI Threat Landscape Report 2025},
  year        = {2025},
  institution = {HiddenLayer, Inc.},
  url         = {https://www.hiddenlayer.com/news/hiddenlayer-ai-threat-landscape-report-reveals-ai-breaches-on-the-rise}
}

@techreport{ms_redteam,
  author      = {{Microsoft Security Response Center}},
  title       = {AI Red Team Building Blocks},
  year        = {2024},
  institution = {Microsoft Corporation},
  url         = {https://learn.microsoft.com/en-us/security/ai-red-teaming}
}

@book{mlsecops_guide,
  title     = {Engineering MLOps: Rapidly build, test, and manage production-ready machine learning life cycles},
  author    = {Raj, Emmanuel},
  year      = {2021},
  publisher = {Packt Publishing}
}

@inproceedings{goodfellow2015,
  author    = {Goodfellow, Ian J. and Shlens, Jonathon and Szegedy, Christian},
  title     = {Explaining and Harnessing Adversarial Examples},
  booktitle = {Proceedings of the International Conference on Learning Representations (ICLR)},
  year      = {2015}
}

@inproceedings{carlini2017,
  author    = {Carlini, Nicholas and Wagner, David},
  title     = {Towards Evaluating the Robustness of Neural Networks},
  booktitle = {Proceedings of the IEEE Symposium on Security and Privacy},
  pages     = {39--57},
  year      = {2017}
}

@inproceedings{greshake2023,
  author    = {Greshake, Kai and Abdelnabi, Sahar and Mishra, Shailesh and Endres, Christoph and Holz, Thorsten and Fritz, Mario},
  title     = {Not What You've Signed Up For: Compromising Real-World LLM-Integrated Applications with Indirect Prompt Injection},
  booktitle = {Proceedings of the ACM Workshop on Artificial Intelligence and Security (AISec)},
  year      = {2023}
}

@misc{iso27005,
  author       = {{ISO/IEC}},
  title        = {ISO/IEC 27005:2022 -- Information security, cybersecurity and privacy protection -- Guidance on managing information security risks},
  year         = {2022},
  howpublished = {International Organization for Standardization}
}

@article{art2018,
  author  = {Nicolae, Maria-Irina and Sinn, Mathieu and Tran, Minh Ngoc and others},
  title   = {Adversarial Robustness Toolbox v1.0.0},
  journal = {arXiv preprint arXiv:1807.01069},
  year    = {2018}
}

@misc{garak2024,
  author       = {{NVIDIA}},
  title        = {garak: LLM Vulnerability Scanner},
  year         = {2024},
  howpublished = {\url{https://github.com/NVIDIA/garak}}
}

\end{document}